\documentclass[letterpaper, 10 pt, conference]{ieeeconf}  

\IEEEoverridecommandlockouts                              

\usepackage{amssymb}  

\usepackage{xcolor}
\usepackage{colortbl,booktabs}
\usepackage{graphicx}
\usepackage{xcolor}
\usepackage{fancyhdr}
\usepackage{tabularx}
\usepackage{multirow}

\usepackage{cite}

\title{\LARGE \bf
Self-supervised Assisted Active Learning for Skin Lesion Segmentation
}

\author{ Ziyuan Zhao$^{*1,2,3}$, Wenjing Lu$^{*1,4}$, Zeng Zeng$^{\dagger 1,2}$, Kaixin Xu$^{1}$, Bharadwaj Veeravalli$^{4}$, Cuntai Guan$^{3}$
\thanks{* Contributed equally. $^{\dagger}$Corresponding author, email: zengz@i2r.a-star.edu.sg. This research is supported by Institute for Infocomm Research (I2R), Agency for Science, Technology and Research (A*STAR), Singapore. $^{1}$ I2R, A*STAR, Singapore. $^{2}$Artificial Intelligence, Analytics And Informatics (AI$^3$), A*STAR, Singapore. $^{3}$ Nanyang Technological University, Singapore. $^{4}$ National University of Singapore, Singapore. This work was done when Wenjing was an intern at I2R, A*STAR.}
}

\begin{document}

\maketitle
\thispagestyle{empty}
\pagestyle{empty}



\begin{abstract}

Label scarcity has been a long-standing issue for biomedical image segmentation, due to high annotation costs and professional requirements.
Recently, active learning~(AL) strategies strive to reduce annotation costs by querying a small portion of data for annotation, receiving much traction in the field of medical imaging. 
However, most of the existing AL methods have to initialize models with some randomly selected samples followed by active selection based on various criteria, such as uncertainty and diversity. 
Such random-start initialization methods inevitably introduce under-value redundant samples and unnecessary annotation costs.
For the purpose of addressing the issue, we propose a novel self-supervised assisted active learning framework in the cold-start setting, in which the segmentation model is first warmed up with self-supervised learning (SSL), and then SSL features are used for sample selection via latent feature clustering without accessing labels.
We assess our proposed methodology on skin lesions segmentation task. Extensive experiments demonstrate that our approach is capable of achieving promising performance with substantial improvements over existing baselines.

\indent \textit{Clinical relevance}— The proposed method can smartly select samples to annotate without requiring labels for model initialization, which can save annotation costs in clinical practice. 

\end{abstract}

\section{INTRODUCTION}
Segmenting lesions and pathological structures from medical images has received great attention in clinical practice. 
Recently, deep learning has set up an incredibly high standard in semantic segmentation tasks~\cite{long2015fully,ronneberger2015u,zhao2021multi}. 
However, such success of deep neural networks always comes with the price of massive amounts of high-accuracy annotations. Additionally, obtaining labeled large-scale datasets at pixel-level is time-consuming, labor-intensive and expensive, requiring extensive clinical experience.

Common label scarcity scenarios encourage research into alleviating the annotation burden under limited supervision, including semi-supervised learning (SSL)~\cite{li2020transformation,zhao2021dsal,li2021hierarchical}, self-supervised learning~\cite{SeSeNet2019, zhou2019models}, transfer learning (TL)~\cite{kouw2018introduction,an2020transfer,zhuang2020comprehensive, zhao2021mt} and active learning (AL)~\cite{gorriz2017cost,shi2019active, zhao2020deeply}.
More recently, deep active learning receives much attention, since it can reduce annotation costs while maintaining model performance.
AL is developed to judiciously query valuable samples for annotation based on various criteria and then update the current model in an iterative manner~\cite{tajbakhsh2020embracing}.
To obtain an initially trained segmentation model, AL requires a minimal set of base annotations for initialization, while most existing AL methods randomly select samples from the unlabeled pool with cold-start~\cite{yang2017suggestive,beluch2018power,kuo2018cost}, followed by active selection based on different criteria, including entropy~\cite{gorriz2017cost}, diversity~\cite{zhou2017fine} and representation~\cite{shi2019active}. However, such cold-start/random-start initialization may introduce under-value labels, leading to annotation redundancy. Moreover, suboptimal initial models may affect the quality of the following AL selection, especially for model-based methods~\cite{gorriz2017cost,shi2019active}.
By enforcing transformation consistency between similar unlabeled samples, self-supervised learning~(SSL) methods can learn transferable representations for downstream tasks~\cite{jaiswal2021survey,zhou2019models,he2020momentum}, which motivates us to explore the robust representation ability of SSL for model initialization and sample selection.

In this work, we introduce a novel warm-start active learning method, in which we first train a self-supervised model to learn deep features in the latent space, and leverage distance-based clustering to build an initial representative set for annotation, thereby reducing annotation costs and minimizing the redundancy.
To validate our proposed methods, extensive experiments are conducted on ISIC~2017 skin lesion dataset~\cite{codella2018skin}.
Quantitative results indicate that our approach exceeds various baselines by large margins, with the proposed sample selection for AL initialization.

\begin{figure*}[htb]
\centering
\includegraphics[width=0.68\textwidth]{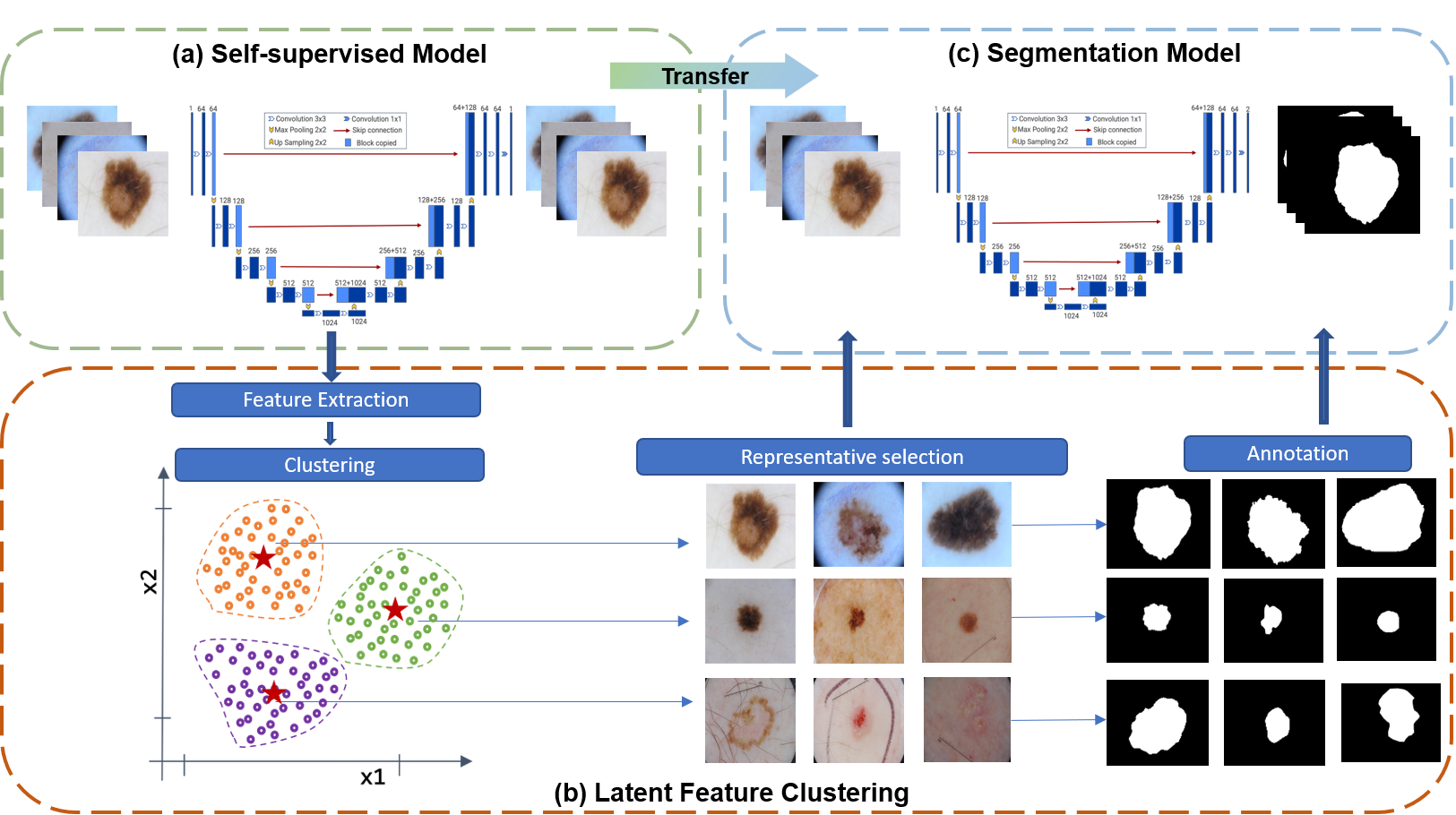}
\caption{Overall pipeline of our method: (a) Self-supervised model: pre-trained reconstructed network as well as feature extraction network; (b) Clustering-based feature extraction: obtain feature descriptors in latent space from the pre-trained self-supervised network and select the most representative sample according to the feature clustering order in proportion; (c) Segmentation Model: query for the selected samples for segmentation task which model is initialized by the pre-trained self-supervised weights.}
\label{fig:pipeline}
\vspace{-10pt}
\end{figure*}

\section{RELATED WORK}

In active learning, the model is fine-tuned iteratively with newly-annotated sampled by selecting unlabeled samples for annotation using various AL strategies based on uncertainty and diversity~\cite{tajbakhsh2020embracing}. With the great progress of deep learning, deep active learning~(DAL) has emerged~\cite{ren2021survey,zhou2017fine,gorriz2017cost,zhao2021dsal}. For instance, Zhou~\emph{et al}~\cite{zhou2017fine} proposed an AIFT framework, in which, the model is continuously fine-tuned, and valuable samples are selected based on model uncertainty and diversity for gradually improving model performance. CEAL~\cite{gorriz2017cost} adopts a complementary sample selection strategy for skin lesion segmentation by assigning pseudo labels to samples with low prediction confidence and querying annotations for those samples with high uncertainty iteratively.
Despite the great success of these methods, most existing DAL methods focus on sample selection using model-based criteria and highly rely on the initial model performance trained with randomly-annotated labels, ignoring the validity and feasibility of initial labels. Zheng~\emph{et al}~\cite{zheng2019biomedical} developed representative annotation (RA), a one-shot active learning method for fast selecting representative samples through extracted features from an unsupervised network.

Recently, self-supervised learning has witnessed the significant potential for learning visual representations, achieving promising performance for a wide range of computer vision applications~\cite{jing2020self}. In self-supervised learning, pretext tasks with unlabeled data are designed for visual representation learning and model initialization. For medical imaging, Zhou~\emph{et al}~\cite{zhou2019models} devised a self-supervised learning framework,~\emph{i.e.}, ModelGenesis, in which, reconstruction loss is constructed with various transformation techniques to improve self-supervised performance for downstream tasks. We are inspired to advance SSL into AL for model warm-up and representative sample selection.

\section{METHODOLOGY}
Fig.~\ref{fig:pipeline} illustrates the proposed active learning architecture, in which, a self-supervised learning framework is firstly built to explore features from unlabeled dataset as well as warm up the segmentation model. Then, based on clustering results of latent features extracted from SSL, we develop a criterion to select representative samples from unlabeled data, forming the initial dataset for training with warm-start. In this self-supervised manner, we make full use of robust representation learning of SSL from both aspects of the model and features for model initialization and sample selection.

\subsection{Self-Supervised Learning Model}
In our framework, U-Net~\cite{ronneberger2015u}, a well-known encoder-decoder architecture, serves as the backbone segmentation network in our system and is also utilized for self-supervised learning. To learn robust features from unlabeled data, we follow ModelGenesis~\cite{zhou2019models} to perform a series of deformation operations and enforce reconstruction consistency between original images and responding transformed ones. Specially, we denote the training dataset as $\mathcal{X}=\left\{\left(x_{i}\right)\right\}_{i=1}^{N}$, where $x_i$ is the input 2D medical image, and $\mathrm{N}$ is the number of samples. The model is optimized by minimizing the reconstruction loss between original images and recovered images as:
\begin{equation}
\min _{\theta} \sum_{i=1}^{N} l\left(f\left(D\left(x_{i}\right) ; \theta\right), x_{i}\right),
\end{equation}

where $l$ is the reconstruction loss, $f(\cdot)$ is the self-supervised neural network, $D(\cdot)$ is the deformation operations, and $\theta$ denotes the model weights. We adopted $L_2$ distance to measure the reconstruction loss, and deformation operations including nonlinear changes, local pixel perturbation out/in-painting were adopted to learn visual features,~\emph{e.g.}, surfaces and texture via self-supervised learning.

\subsection{Embedded Feature Clustering}
The latent features of the last encoder are always transferred for downstream tasks, such as classification. Following the paradigm of SSL, similar images should have similar latent features. In other words, the neighbors in latent space to the inferred cluster centroids should be more representative. To cluster SSL features in a compact space, we first adopt adaptive average pooling (AAP)~\cite{van2019evolutionary} to downsample the SSL features extracted from the last encoder layer for reducing dimensions and filtering some redundant information. Then, we apply $K$-means~\cite{macqueen1967some, shen2021semi} for clustering, and its objective can be expressed as:
%
\begin{equation}
\mathop{\arg\min}_{\mathbf{S}} \sum_{i=1}^{k} \sum_{\mathbf{x}\in S_{i}}\left\|\phi\left(\mathbf{x}_{\mathbf{i}}\right)-\mu_{i}\right\|^{2},
\end{equation}
where $k$ is the number of clusters, and $S_i$ is the cluster subset, ${\mu}_{i}$ is the mean of all points in $S_i$. The centroids can be regarded as the representative of the cluster, and then we select the centroids and their $N_{C_i}$ neighbors based on $L_2$ distance to form our initial dataset.
The number of selected samples in cluster $C_i$ is defined as $N_{C_{i}}=\left[{C} \times N_{i} / {N}\right]$, where $C$ is the number of candidates for annotation to build the initial training dataset, $N_{i}$ is the number of samples in cluster $C_i$, and the total number of samples is denoted by {N} in the dataset. In this manner, the representative samples were selected to warm up the model.



\subsection{Segmentation Model}
Prior to the AL procedure, the pre-trained weights from the self-supervised learning model are adopted to initialize the segmentation model. The segmentation loss in our model is formatted as 
\begin{equation}
L_{seg}=L_{ce}+ L_{dice},
\end{equation}
where $L_{ce}$ is the cross-entropy loss, and $L_{dice}$ is the Dice loss, which is widely used in various segmentation tasks~\cite{patravali20172d}. Similar to other AL methods, our model is further finetuned with the enlarged labeled dataset iteratively based on our representative criterion. Alternatively, our model can be trained with the initially labeled dataset in a one-stop manner. 


\begin{table}[thb]
\centering
\caption{Dice coefficient of different segmentation methods on ISIC~2017 dataset initializing w/ or w/o pre-trained weights.}
\label{tab:results}
\scalebox{0.95}{
\begin{tabular}{c|c|c|c|c} 
\hline
Select Criteria          & \multicolumn{2}{c|}{Random} & \multicolumn{2}{c}{Our method}  \\ 
\hline
\multirow{2}{*}{dataset} & \multicolumn{4}{c}{Pre-trained weights}                     \\ 
\cline{2-5}
                         & W/o & W/       & W/o & W/           \\ 
\hline
300                      & $0.580$ & $0.682$           & $0.744$ & $0.732$               \\
370                      & $0.657$ & $0.576$           & $0.683$ & $0.682$               \\
440                      & $0.717$ & $0.696$           & $0.743$ & $0.763$               \\
510                      & $0.687$ & $0.725$           & $0.765$ & $0.777$               \\
580                      & $0.774$ & $0.785$           & $0.797$ & $0.803$               \\
\hline
\end{tabular}
}
\end{table}

\section{EXPERIMENTS}
\subsection{Dataset and Experimental Settings}
The ISIC 2017 dataset~\cite{codella2018skin}, which contains $2000$ RGB dermoscopy images manually annotated by medical specialists, was chosen for our experimental assessment. For preprocessing, the dermoscopy images were resized to $256 \times 256$ and transformed to grayscale, followed by min-max normalization. We formed the unlabeled pool of $1,600$ unlabeled images for representative selection and another $400$ for testing. The most common segmentation metric, Dice coefficient~\cite{zhao2021dsal} was used to evaluate the segmentation performance, which calculates the overlap ratio between ground truth labels and predicted segmentation masks. Thus, a higher Dice would mean better segmentation performance.

According to the representative rank-based selective criteria via clustering, $300$ samples were selected for annotation and formed as the base training set. Following~\cite{gorriz2017cost}, we set initial training epochs as $10$, the number of iteration $T = 9$, and at each iteration $t$, $35$ labeled samples were added to the training set for training $2$ epochs.
In our method, Adaptive Average Pooling (AAP) was used to reduce dimensionality to process the SSL features from a dimension of $512\times16\times16$ to $512\times2\times2$, obtaining $2048-$dimensional features after flattening. The number of clusters here was set to $10$.

\begin{figure}[!htb]
\centering
\includegraphics[width=0.4\textwidth]{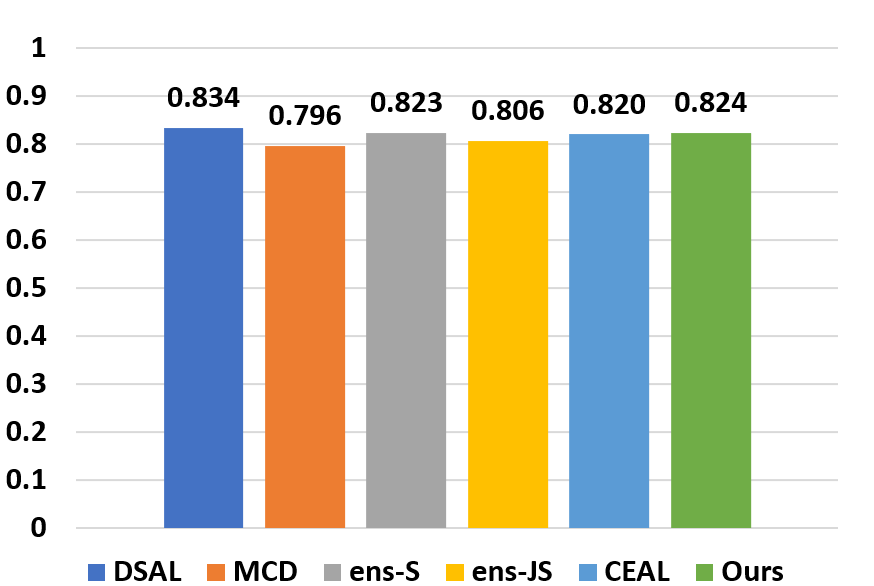}
\caption{Comparison on Dice coefficient of different AL methods.}
\label{fig:results}
\end{figure}
\subsection{Results and Discussions}


To corroborate the efficiency of our initial sample selection strategy, we first compare the proposed approach with a random query strategy on base set initialization and following a few iterations. Table~\ref{tab:results} summarizes the experimental results. We can observe that initialization with samples based on random query achieved $58\%$ in Dice, while our AL criteria can achieve much better performance, which demonstrates that it is feasible to select representative samples based on SSL features for annotation. We further implemented both the random query strategy and our method with pre-trained SSL weights, and observe that pre-trained SSL weights can boost the segmentation performance. Through the first few iterations, the training set was enlarged to $580$. A considerable improvement in segmentation performance over the baseline random query method. It is noted that the pre-trained weights may decrease performance slightly for the initial few iterations, especially for random queries, which possibly because the pre-trained SSL features take the dominant of the model training with only few training epochs. Along with the iterations increase, more task-related features are learned, boosting the segmentation performance.

To further evaluate the effectiveness of the proposed method for AL, we compared our approach with representative AL counterparts, including ensemble-S (ens-S)~\cite{yang2017suggestive,beluch2018power}, ensemble-JS~(ens-JS)~\cite{kuo2018cost}, MC Dropout~(MCD)~\cite{gal2017deep}, CEAL~\cite{gorriz2017cost}, and DSAL~\cite{zhao2021dsal} on the same AL scenario~\cite{zhao2021dsal}. The final results after $9$ iterations are shown in Fig.~\ref{fig:results}. It can be observed that the proposed method achieved comparable performance among different AL methods, which demonstrates that our method can not only help initial set selection but also serve as an AL criterion for iterative training.

To investigate the influence of cluster number and dimensionality, we squeeze features from a dimension of $512\times16\times16$ to $512\times2\times2$ or $512\times1\times1$, obtaining $2048-$dimensional or $512-$dimensional features after flattening, while the numbers of clusters were set to $10$ or $5$ as shown in Table~\ref{mytable_model}. We observe that increasing the feature dimensions and the number of clusters can help improve the model performance, sicne higher dimensions of features can preserve more important spatial information, while more clustering centroids can help reduce information redundancy and extract more representative samples in the feature space.



\section{CONCLUSIONS}
This paper presents a deep feature selection strategy based on self-supervised learning by projecting into the hidden space to query representative samples, and the initialization strategy of AL with pre-trained weights from self-supervised learning. 
According to extensive experiments, the proposed AL scheme successfully boosts the baseline performance by a significant margin. Furthermore, experiments with different clusters and feature dimensions have been conducted to verify the efficiency of the various components of our framework. In the future, combinations with other selection criteria could be explored, and we could implement our methods on more medical segmentation datasets to evaluate the generalization ability.

\begin{table}[!thb]
\centering
\setlength\tabcolsep{2.5pt}
\caption{Segmentation results of different methods on ISIC~2017 dataset without initializing pre-trained weights.}
\label{mytable_model}
\scalebox{0.95}{
\begin{tabular}{c|c|c|c|c|c}
\hline \multirow{2}{*}{ iter } & \multirow{2}{*}{ Random } & \multicolumn{2}{|c|}{ Cluster 5 } & \multicolumn{2}{c}{ Cluster 10 } \\
\cline { 3 - 6 } & & AAP 512 & AAP 2048 & AAP 512 & AAP 2048 \\
\hline base & $0.580$ & $0.714$ & $0.734$ & $0.739$ & $\mathbf{0 . 7 4 4}$ \\
$t=1$ & $0.614$ & $\mathbf{0 . 6 7 6}$ & $0.673$ & $0.667$ & $0.647$ \\
$t=2$ & $0.657$ & $0.663$ & $\mathbf{0 . 6 9 0}$ & $0.682$ & $0.683$ \\
$t=3$ & $0.645$ & $0.723$ & $0.720$ & $0.699$ & $\mathbf{0 . 7 1 6}$ \\
$t=4$ & $0.717$ & $0.737$ & $0.719$ & $0.736$ & $\mathbf{0 . 7 4 3}$ \\
$t=5$ & $0.708$ & $0.749$ & $\mathbf{0 . 7 5 3}$ & $0.741$ & $0.750$ \\
$t=6$ & $0.687$ & $\mathbf{0 . 7 7 2}$ & $0.767$ & $0.741$ & $0.765$ \\
$t=7$ & $0.763$ & $0.762$ & $0.762$ & $0.747$ & $\mathbf{0 . 7 6 7}$ \\
$t=8$ & $0.774$ & $0.786$ & $0.784$ & $0.779$ & $\mathbf{0 . 7 9 7}$ \\
$t=9$ & $0.766$ & $0.793$ & $0.793$ & $0.776$ & $\mathbf{0 . 8 0 0}$ \\
\hline
\end{tabular}
}
\end{table}

\bibliographystyle{IEEEbib}
\bibliography{refs1.bib}

\begin{thebibliography}{10}

\bibitem{long2015fully}
Jonathan Long, Evan Shelhamer, and Trevor Darrell,
\newblock ``Fully convolutional networks for semantic segmentation,''
\newblock in {\em Proceedings of the IEEE conference on computer vision and
  pattern recognition}, 2015.

\bibitem{ronneberger2015u}
Olaf Ronneberger, Philipp Fischer, and Thomas Brox,
\newblock ``U-net: Convolutional networks for biomedical image segmentation,''
\newblock in {\em International Conference on Medical image computing and
  computer-assisted intervention}. Springer, 2015.

\bibitem{zhao2021multi}
Ziyuan Zhao, Zeyu Ma, Yanjie Liu, Zeng Zeng, and Pierce~KH Chow,
\newblock ``Multi-slice dense-sparse learning for efficient liver and tumor
  segmentation,''
\newblock in {\em EMBC}. IEEE, 2021.

\bibitem{li2020transformation}
Xiaomeng Li, Lequan Yu, Hao Chen, Chi-Wing Fu, Lei Xing, and Pheng-Ann Heng,
\newblock ``Transformation-consistent self-ensembling model for semisupervised
  medical image segmentation,''
\newblock {\em IEEE Transactions on Neural Networks and Learning Systems},
  2020.

\bibitem{zhao2021dsal}
Ziyuan Zhao, Zeng Zeng, Kaixin Xu, Cen Chen, and Cuntai Guan,
\newblock ``Dsal: Deeply supervised active learning from strong and weak
  labelers for biomedical image segmentation,''
\newblock {\em IEEE Journal of Biomedical and Health Informatics}, 2021.

\bibitem{li2021hierarchical}
Shumeng Li, Ziyuan Zhao, Kaixin Xu, Zeng Zeng, and Cuntai Guan,
\newblock ``Hierarchical consistency regularized mean teacher for
  semi-supervised 3d left atrium segmentation,''
\newblock in {\em 2021 43rd Annual International Conference of the IEEE
  Engineering in Medicine \& Biology Society (EMBC)}. IEEE, 2021.

\bibitem{SeSeNet2019}
Zeng Zeng, Xulei Yang, Yu~Qiyun, Yao Meng, and Zhang Le,
\newblock ``Sese-net: Self-supervised deep learning for segmentation,''
\newblock {\em Pattern Recognition Letters}, 2019.

\bibitem{zhou2019models}
Zongwei Zhou, Vatsal Sodha, Md~Mahfuzur Rahman~Siddiquee, Ruibin Feng, Nima
  Tajbakhsh, Michael~B. Gotway, and Jianming Liang,
\newblock ``Models genesis: Generic autodidactic models for 3d medical image
  analysis,''
\newblock in {\em International Conference on Medical Image Computing and
  Computer-Assisted Intervention}. Springer, 2019.

\bibitem{kouw2018introduction}
Wouter~M Kouw and Marco Loog,
\newblock ``An introduction to domain adaptation and transfer learning,''
\newblock {\em arXiv:1812.11806}, 2018.

\bibitem{an2020transfer}
Sizhe An, Ganapati Bhat, Suat Gumussoy, and Umit Ogras,
\newblock ``Transfer learning for human activity recognition using
  representational analysis of neural networks,''
\newblock {\em arXiv preprint arXiv:2012.04479}, 2020.

\bibitem{zhuang2020comprehensive}
Fuzhen Zhuang, Zhiyuan Qi, Keyu Duan, Dongbo Xi, Yongchun Zhu, Hengshu Zhu, Hui
  Xiong, and Qing He,
\newblock ``A comprehensive survey on transfer learning,''
\newblock {\em Proceedings of the IEEE}, 2021.

\bibitem{zhao2021mt}
Ziyuan Zhao, Kaixin Xu, Shumeng Li, Zeng Zeng, and Cuntai Guan,
\newblock ``Mt-uda: Towards unsupervised cross-modality medical image
  segmentation with limited source labels,''
\newblock in {\em International Conference on Medical Image Computing and
  Computer-Assisted Intervention}. Springer, 2021, pp. 293--303.

\bibitem{gorriz2017cost}
Marc Gorriz, Axel Carlier, Emmanuel Faure, and Xavier Giro-i Nieto,
\newblock ``Cost-effective active learning for melanoma segmentation,''
\newblock {\em arXiv preprint arXiv:1711.09168}, 2017.

\bibitem{shi2019active}
Xueying Shi, Qi~Dou, Cheng Xue, Jing Qin, Hao Chen, and Pheng-Ann Heng,
\newblock ``An active learning approach for reducing annotation cost in skin
  lesion analysis,''
\newblock in {\em International Workshop on Machine Learning in Medical
  Imaging}. Springer, 2019.

\bibitem{zhao2020deeply}
Ziyuan Zhao, Xiaoyan Yang, Bharadwaj Veeravalli, and Zeng Zeng,
\newblock ``Deeply supervised active learning for finger bones segmentation,''
\newblock in {\em 2020 42nd Annual International Conference of the IEEE
  Engineering in Medicine \& Biology Society (EMBC)}. IEEE, 2020.

\bibitem{tajbakhsh2020embracing}
Nima Tajbakhsh, Laura Jeyaseelan, Qian Li, Jeffrey~N Chiang, Zhihao Wu, and
  Xiaowei Ding,
\newblock ``Embracing imperfect datasets: A review of deep learning solutions
  for medical image segmentation,''
\newblock {\em Medical Image Analysis}, 2020.

\bibitem{yang2017suggestive}
Lin Yang, Yizhe Zhang, Jianxu Chen, Siyuan Zhang, and Danny~Z Chen,
\newblock ``Suggestive annotation: A deep active learning framework for
  biomedical image segmentation,''
\newblock in {\em MICCAI}, 2017.

\bibitem{beluch2018power}
William~H Beluch, Tim Genewein, Andreas N{\"u}rnberger, and Jan~M K{\"o}hler,
\newblock ``The power of ensembles for active learning in image
  classification,''
\newblock in {\em Proceedings of the IEEE Conference on Computer Vision and
  Pattern Recognition}, 2018.

\bibitem{kuo2018cost}
Weicheng Kuo, Christian H{\"a}ne, Esther Yuh, Pratik Mukherjee, and Jitendra
  Malik,
\newblock ``Cost-sensitive active learning for intracranial hemorrhage
  detection,''
\newblock in {\em MICCAI}. Springer, 2018.

\bibitem{zhou2017fine}
Zongwei Zhou, Jae Shin, Lei Zhang, Suryakanth Gurudu, Michael Gotway, and
  Jianming Liang,
\newblock ``Fine-tuning convolutional neural networks for biomedical image
  analysis: actively and incrementally,''
\newblock in {\em IEEE CVPR}, 2017.

\bibitem{jaiswal2021survey}
Ashish Jaiswal, Ashwin~Ramesh Babu, Mohammad~Zaki Zadeh, Debapriya Banerjee,
  and Fillia Makedon,
\newblock ``A survey on contrastive self-supervised learning,''
\newblock {\em Technologies}, 2021.

\bibitem{he2020momentum}
Kaiming He, Haoqi Fan, Yuxin Wu, Saining Xie, and Ross Girshick,
\newblock ``Momentum contrast for unsupervised visual representation
  learning,''
\newblock in {\em IEEE CVPR}, 2020.

\bibitem{codella2018skin}
Noel~CF Codella, David Gutman, M~Emre Celebi, Brian Helba, Michael~A Marchetti,
  Stephen~W Dusza, Aadi Kalloo, Konstantinos Liopyris, Nabin Mishra, Harald
  Kittler, et~al.,
\newblock ``Skin lesion analysis toward melanoma detection: A challenge at the
  2017 international symposium on biomedical imaging (isbi), hosted by the
  international skin imaging collaboration (isic),''
\newblock in {\em ISBI 2018}. IEEE, 2018.

\bibitem{ren2021survey}
Pengzhen Ren, Yun Xiao, Xiaojun Chang, Po-Yao Huang, Zhihui Li, Brij~B Gupta,
  Xiaojiang Chen, and Xin Wang,
\newblock ``A survey of deep active learning,''
\newblock {\em ACM Computing Surveys (CSUR)}, 2021.

\bibitem{zheng2019biomedical}
Hao Zheng, Lin Yang, Jianxu Chen, Jun Han, Yizhe Zhang, Peixian Liang, Zhuo
  Zhao, Chaoli Wang, and Danny~Z Chen,
\newblock ``Biomedical image segmentation via representative annotation,''
\newblock in {\em Proceedings of the AAAI Conference on Artificial
  Intelligence}, 2019.

\bibitem{jing2020self}
Longlong Jing and Yingli Tian,
\newblock ``Self-supervised visual feature learning with deep neural networks:
  A survey,''
\newblock {\em IEEE transactions on pattern analysis and machine intelligence},
  2020.

\bibitem{van2019evolutionary}
Gerard~Jacques van Wyk and Anna~Sergeevna Bosman,
\newblock ``Evolutionary neural architecture search for image restoration,''
\newblock in {\em IJCNN}. IEEE, 2019.

\bibitem{macqueen1967some}
James MacQueen et~al.,
\newblock ``Some methods for classification and analysis of multivariate
  observations,''
\newblock in {\em Proceedings of the fifth Berkeley symposium on mathematical
  statistics and probability}, 1967.

\bibitem{shen2021semi}
Xiang Shen, Yinge Sun, Yao Zhang, and Mani Najmabadi,
\newblock ``Semi-supervised intent discovery with contrastive learning,''
\newblock in {\em Proceedings of the 3rd Workshop on Natural Language
  Processing for Conversational AI}, 2021.

\bibitem{patravali20172d}
Jay Patravali, Shubham Jain, and Sasank Chilamkurthy,
\newblock ``2d-3d fully convolutional neural networks for cardiac mr
  segmentation,''
\newblock in {\em International Workshop on Statistical Atlases and
  Computational Models of the Heart}. Springer, 2017.

\bibitem{gal2017deep}
Yarin Gal, Riashat Islam, and Zoubin Ghahramani,
\newblock ``Deep bayesian active learning with image data,''
\newblock in {\em International Conference on Machine Learning}. PMLR, 2017.

\end{thebibliography}

\end{document}